\def\year{2019}\relax
\documentclass[letterpaper]{article} 
\usepackage{aaai19}  
\usepackage{times}  
\usepackage{helvet}  
\usepackage{courier}  
\usepackage{url}  
\usepackage{graphicx}  
\usepackage{amsmath}
\usepackage{amssymb}
\usepackage{algorithm}
\usepackage{algpseudocode}
\usepackage{bm}
\usepackage{booktabs}
\usepackage{multirow}
\usepackage{bbm}
\providecommand{\tightlist}{%
  \setlength{\itemsep}{0pt}\setlength{\parskip}{0pt}}

\DeclareMathOperator\logit{logit}

\def\R{\mathbf{R}}
\def\N{\mathcal{N}}

\def\logitp{\logit p_{ij}}
\def\x{\bm{x}}
\def\one{\mathbbm{1}}

\begin{document}
%
\title{Knowledge Tracing Machines: Factorization Machines for Knowledge Tracing}
\author{Jill-J\^enn Vie\\
RIKEN Center for Advanced Intelligence Project, Tokyo\\
\texttt{vie@jill-jenn.net}
\And Hisashi Kashima\\
Kyoto University \& RIKEN AIP\\
\url{kashima@i.kyoto-u.ac.jp}}

\maketitle

\begin{abstract}
Knowledge tracing is a sequence prediction problem where the goal is to predict the outcomes of students over questions as they are interacting with a learning platform. By tracking the evolution of the knowledge of some student, one can optimize instruction.
Existing methods are either based on temporal latent variable models, or factor analysis with temporal features.
We here show that factorization machines (FMs), a model for regression or classification, encompasses several existing models in the educational literature as special cases, notably additive factor model, performance factor model, and multidimensional item response theory. We show, using several real datasets of tens of thousands of users and items, that FMs can estimate student knowledge accurately and fast even when student data is sparsely observed, and handle side information such as multiple knowledge components and number of attempts at item or skill level. Our approach allows to fit student models of higher dimension than existing models, and provides a testbed to try new combinations of features in order to improve existing models.
\end{abstract}

Modeling student learning is key to be able to detect students that need
further attention, or recommend automatically relevant learning
resources. Initially, models were developed for students sitting for
standardized tests, where students could read every problem statement,
and missing answers could be treated as incorrect. However, in online
platforms such as MOOCs, students attempt some exercises, but do not
even look at other ones. Also, they may learn between different
attempts. How to measure knowledge when students have attempted
different questions?

We want to predict the performance of a set \(I\) of students, say
users, over a set \(J\) of questions, say items (we will interchangeably
refer to questions as items, problems, or tasks). Each student can
attempt a question multiple times, and may learn between successive
attempts. We assume we observe ordered triplets
\((i, j, o) \in I \times J \times \{0, 1\}\) which encode the fact that
student \(i\) attempted question \(j\) and got it either correct
(\(o = 1\)) or incorrect (\(o = 0\)). Triplets are sorted
chronologically. Then, given a new pair \((i', j')\), we need to predict
whether student \(i'\) will get question \(j'\) correct or incorrect. We
can also assume extra knowledge about users, or items.

So far, various models have been designed for student modeling, either
based on prediction of sequences \cite{piech2015deep}, or factor
analysis \cite{thai2011factorization,lavoue2018adaptive}. Most of
existing techniques model students or questions with unidimensional
parameters. In this paper, we generalize these models to higher
dimensions and manage to train efficiently student models of dimension
up to 20. Our family of models is particularly convenient when
observations from students are sparse, e.g.~when some students attempted
few questions, or some questions were answered by few students, which is
most of the data usually encountered in online platforms such as MOOCs.

When fitting student models, it is better to rely on all the information
available at hand. In order to get information about questions, one can
identify the knowledge components (KCs) involved in each question. This
side information is usually encoded under the form of a \emph{q-matrix},
that maps items to knowledge components: \(q_{jk}\) is 1 if item \(j\)
involves KC \(k\), 0 otherwise. In this paper, we will also note
\(KC(j)\) the sets of skills involved by question \(j\),
i.e.~\(KC(j) = \{k|q_{jk} = 1\}\).

In order to model different attempts, one can keep track of how many
times a student has attempted a question, or how many times a student
has had the opportunity to acquire a skill, while interacting with the
learning material.

Our experiments show, in particular, that:

\begin{itemize}
\tightlist
\item
  It is better to estimate a bias for each item (not only skill), which
  popular educational data mining (EDM) models do not.
\item
  Most existing models in EDM cannot handle side information such as
  multiple skills for one item, but the proposed approach does.
\item
  Side information improves performance more than increasing the latent
  dimension.
\end{itemize}

To the best of our knowledge, this is the most generic framework that
incorporates side information into a student model. For the sake of
reproducibility, our implementation is available on
GitHub\footnote{\url{https://github.com/jilljenn/ktm}}. The interested
reader can check our code and reuse it in order to try new combinations
and devise new models.

In Section 2, we show related work. In Section 3, we present a family of
models, knowledge tracing machines, and recover famous models of the EDM
literature as special cases. Then, in Section 4 we conduct experiments
and show our results in Section 5. We conclude with further work in
Section 6.

\section{Related Work}

In this section, we review several approaches proposed to model student
learning.

\subsection{Knowledge Tracing}

Knowledge Tracing aims at predicting the sequence of outcomes of a
student over questions. It usually relies on modeling the state of the
learner throughout the process. After several attempts, students may
eventually evolve to a state of mastery.

The most popular model is Bayesian knowledge tracing (BKT), which is a
hidden Markov model \cite{corbett1994knowledge}. However, BKT cannot
model the fact that a question might require several KCs. New models
have been proposed that do handle multiple subskills, such as
feature-aware student tracing (FAST) \cite{gonzalez2014general}.

As deep learning models have proven successful at predicting sequences,
they have been applied to student modeling: deep knowledge tracing (DKT)
is a long short-term memory (LSTM) \cite{piech2015deep}. Several
researchers have reproduced the experiment on several variations of the
Assistments dataset
\cite{xiong2016going,wilson2016back,wilson2016estimating}, and shown
that some factor analysis models could match the performance of DKT, as
we will see now.

\subsection{Factor Analysis}

Factor analysis tend to learn common factors in data in order to
generalize observations. They have been successfully applied to matrix
completion, where we assume that data is recorded for (user, item)
pairs, but many entries are missing. The main difference with sequence
prediction for our purposes is that the order in which the data is
observed does not matter. If one wants to encode temporality though, it
is possible to complement the data with temporal features such as simple
counters, as we will see later. In all that follows, \(\logit\) will
denote the logit function: \(\logit p = \log \frac{p}{1 - p}\).

\subsubsection{Item Response Theory}

The most simple model for factor analysis does not assume knowledge
between several attempts, it is the 1-parameter logistic item response
theory model, also known as Rasch model:

\[ \logitp = \theta_i - d_j \]

\noindent where \(\theta_i\) measures the ability of student \(i\) (the
student bias) and \(d_j\) measures the difficulty of question \(j\) (the
question bias). We will refer to the Rasch model as IRT in the rest of
the paper. More recently, Wilson have shown that IRT could outperform
DKT \cite{wilson2016estimating}, even without temporal features
\cite{gonzalez2014general}. It may be because DKT have many parameters
to estimate, so they are prone to overfitting, and they are hard to
train on long sequences.

The IRT model has been extended to multidimensional abilities:

\[ \logitp = \langle \bm{\theta_i}, \bm{d_j} \rangle + \delta_j \]

\noindent where \(\delta_j\) is the easiness of item \(j\) (item bias).
Multidimensional Item Response Theory (MIRT) has a reputation to be hard
to train \cite{desmarais2012review} thus is not frequently encountered
in the EDM literature, and still, the dimensionality used in
psychometrics papers is up to 4, but we show in this paper how to train
those models effectively, up to 20 dimensions.

\subsubsection{AFM and PFA}

Additive factor model (AFM) \cite{cen2006learning} takes into account
the number of attempts a learner has made to an item:

\[ \logitp = \sum_{k \in KC(j)} \beta_k + \gamma_k N_{ik} \]

\noindent where \(\beta_k\) is the bias for the skill \(k\), and
\(\gamma_k\) the bias for each opportunity of learning the skill \(k\).
\(N_{ik}\) is the number of attempts of student \(i\) over a question
that requires the skill \(k\).

Performance factor analysis (PFA) \cite{pavlik2009performance} counts
separately positive and negative attempts:

\[ \logitp = \sum_{k \in KC(j)} \beta_k + \gamma_k W_{ik} + \delta_k F_{ik} \]

\noindent where \(\beta_k\) is the bias for the skill \(k\),
\(\gamma_k\) (\(\delta_k\)) the bias for each opportunity of learning
the skill \(k\) after a successful (unsuccessful) attempt, \(W_{ik}\)
(\(F_{ik}\)) is the number of successes (failures) of student \(i\) over
a question that requires the skill \(k\). In other words, AFM can be
seen as a particular case of PFA where \(\gamma_k = \delta_k\) for every
skill \(k\). Please note that AFM and PFA do not consider item
difficulty, presumably to avoid the item cold-start problem. According
to \cite{gonzalez2014general}, PFA and FAST have comparable performance.
By reproducing experiments, \cite{xiong2016going} have managed to match
the performance of DKT with PFA.

\subsection{Factorization machines}

Numerous works have coined the similarity between student modeling and
collaborative filtering (CF) in recommender systems
\cite{bergner2012model,thai2011factorization}. For CF, factorization
machines were designed to provide a way to encode side information about
items or users into the model.

\cite{thai2012using} and \cite{sweeney2016next} have used factorization
machines in their regression form for student modeling (where the root
mean squared error is used as metric) but to the best of our knowledge,
it has not been used in its classification form for student modeling.
This is what we present in this paper, in the next section.

\section{Knowledge tracing machines}

\begin{table*}
\centering
\caption{An example of encoding for training a Knowledge Tracing Machine.}
\begin{tabular}{cc c ccc c ccc c ccc c ccc }
\toprule
  \multicolumn{2}{c}{Users}  & & \multicolumn{3}{c}{Items} & & \multicolumn{3}{c}{Skills}  & & \multicolumn{3}{c}{Wins}  & &  \multicolumn{3}{c}{Fails}  \\
\cmidrule{1-2}
\cmidrule{4-6}
\cmidrule{8-10}
\cmidrule{12-14}
\cmidrule{16-18}
  1 & 2 & & Q$_1$ & Q$_2$ & Q$_3$ & & KC$_1$ & KC$_2$ & KC$_3$ & & KC$_1$ & KC$_2$ & KC$_3$ & & KC$_1$ & KC$_2$ & KC$_3$\\
\midrule
  0 &   1 & &  0 &   1 &   0 &&   1 &   1 &   0 &&   0 &   0 &   0 &&   0 &   0 &   0 \\
  0 &   1 & &  0 &   1 &   0 &&   1 &   1 &   0 &&   1 &   1 &   0 &&   0 &   0 &   0 \\
  0 &   1 & &  0 &   1 &   0 &&   1 &   1 &   0 &&   1 &   1 &   0 &&   1 &   1 &   0 \\
  0 &   1 & &  0 &   0 &   1 &&   0 &   1 &   1 &&   0 &   2 &   0 &&   0 &   1 &   0 \\
  0 &   1 & &  0 &   0 &   1 &&   0 &   1 &   1 &&   0 &   2 &   0 &&   0 &   2 &   1 \\
  1 &   0 & &  0 &   1 &   0 &&   1 &   1 &   0 &&   0 &   0 &   0 &&   0 &   0 &   0 \\
  1 &   0 & &  1 &   0 &   0 &&   0 &   0 &   0 &&   0 &   0 &   0 &&   0 &   0 &   0 \\
\bottomrule
\end{tabular}
\hspace{1mm}
\begin{tabular}{c}
\toprule
\multirow{2}[3]{*}{Outcome} \\[6mm]
\midrule
  1\\0\\1\\0\\1\\1\\0\\
\bottomrule
\end{tabular}

\label{encoding-features}
\end{table*}

We now introduce the family of models described in this paper, Knowledge
Tracing Machines (KTM).

Let \(N\) be the number of features. Features can refer either to
students, exercises, knowledge components (KCs), opportunities for
learning, or extra information about the learning environment. For
example, one might want to model the fact that the student attempted an
exercise on mobile, or on computer, which might influence their outcome:
it may be harder to type a correct answer when using a mobile, so this
data should be taken into account in the predictions.

KTMs model the probability of observing binary outcomes of events (right
or wrong), based on a sparse set of weights for all features involved in
the event. Features involved in an event are encoded by a sparse vector
\(\x\) of length \(N\) such that \(x_i > 0\) iff feature
\(1 \leq i \leq N\) is involved in the event. For each event involving
\(\x\), the probability \(p(\x)\) to observe a positive outcome
verifies:

\begin{equation}
\psi(p(\x)) = \mu + \underbrace{\sum_{k = 1}^N w_k x_k}_{\textnormal{logistic regression}} + \underbrace{\sum_{1 \leq k < l \leq N} x_k x_l \langle \bm{v_k}, \bm{v_l} \rangle}_{\textnormal{pairwise interactions}}
\label{ktm}
\end{equation}

\begin{figure}[b]
\includegraphics[width=\linewidth]{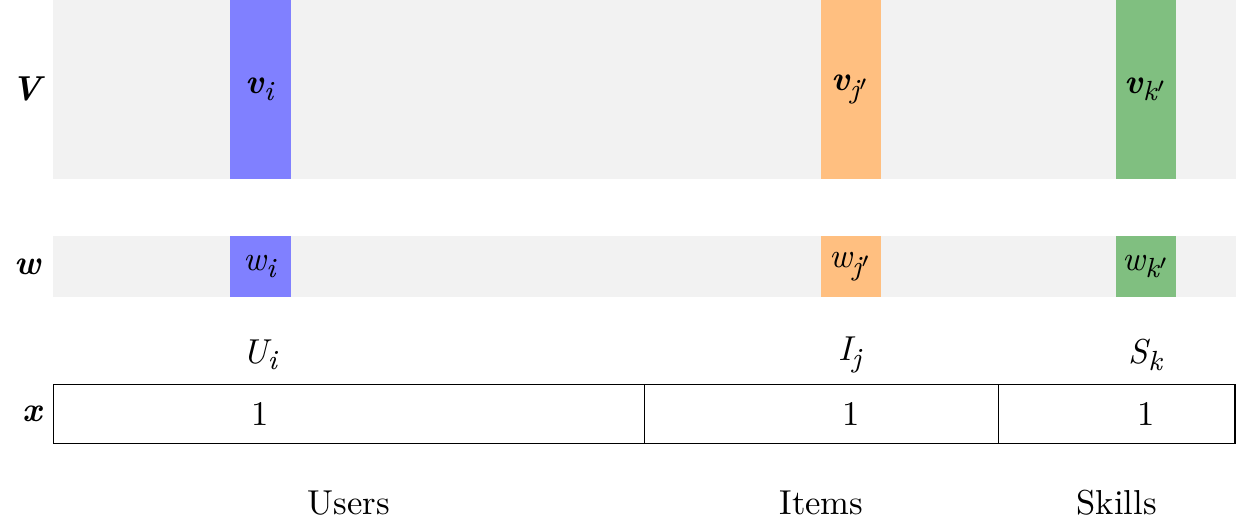}
\caption{Example of activation of a knowledge tracing machine.}
\label{fm-activation}
\end{figure}

\noindent where \(\psi\) is a link function such as \(\logit\), \(\mu\)
is a global bias, each feature \(i\) is modeled by both a bias
\(w_i \in \R\) and an embedding \(\bm{v_i} \in \R^d\) for some dimension
\(d\). In what follows, \(\bm{w}\) will refer to the vector of biases
\((w_1, \ldots, w_N)\) and \(\bm{V}\) to the matrix of embeddings
\(\bm{v_i}\), \(i = 1, \ldots, N\). For each event, only the features
that have \(x_i > 0\) will contribute to the prediction, see
Figure~\ref{fm-activation}.

\subsection{Data and encoding of side information}

We now describe how to encode the observed data in the learning platform
into the sparse vector \(\x\). First, we need to choose which features
will be represented in the modeling.

\subsubsection{Users}

Let us assume there are \(n\) students. The first \(n\) features will be
for all \(n\) students. As an example, if student \(1 \leq i \leq n\) is
involved in the observation, its \(x_i\) value will be set to 1, while
the ones for the other students will be set to 0. This is called a
one-hot vector.

\subsubsection{Items}

Let us assume there are \(m\) questions or items. One can allocate \(m\)
more features for all \(m\) questions. If question \(1 \leq j \leq m\)
is involved in the observation, its component in \(\x\) will be set to
1, while the ones for the other questions will be set to 0.

\subsubsection{Skills}

We now assume there are \(s\) skills. We can then allocate \(s\) extra
features for those \(s\) skills. The skills involved in an observation
of a student over a question \(j\) are the ones of \(KC(j)\).

\subsubsection{Attempts}

One can allocate \(s\) extra features as counters of how many
opportunities a student could have learned a skill involved in the test.

\subsubsection{Wins \& Fails}

One can also distinguish between successes and failures: allocate \(s\)
features as opportunities to have learned a skill if the attempt was
correct, \(s\) more features as opportunities to have learned a skill if
the attempt was incorrect.

\subsubsection{Extra side information}

More side information can be concatenated to the existing sparse
features, such as the school ID and teacher ID of the student, or also
other information such as the type of test: low stake (practice) or high
stake (posttest), etc.

\subsubsection{Full example}

See Table~\ref{encoding-features} for an example of encoding of users +
items + skills + wins + fails, for the set of observed, chronologically
ordered triplets \((2, 2, 1)\) (student 2 attempted question 2 and got
it correct), \((2, 2, 0)\), \((2, 2, 1)\), \((2, 3, 0)\), \((2, 3, 1)\),
\((1, 2, 1)\), \((1, 1, 0)\). Here, we assume that there are \(n = 2\)
students, \(m = 3\) questions, \(m = 3\) skills and question 1 does not
involve any skill, question 2 involves skills 1 and 2, question 3
involves skills 2 and 3. At the beginning, user 2 had no opportunity to
learn any skill, so counters of wins and fails are all 0. After student
2 got question 2 correct, as it involved skills 1 and 2, the counters of
wins for these two skills are incremented, and encoded for the next
observation. We thus managed to encode the triplets with
\(N = n + m + 3s = 14\) features, and at training time, a bias and an
embedding will be learned for each one of them.

\begin{table*}[t]
\centering
\caption{Datasets used for the experiments}
\begin{tabular}{lrrrrrrr}
\toprule
        Name &  Users &  Items &  Skills &  Skills per item &  Entries &  Sparsity (user, item) &  Attempts per user \\
\midrule
    fraction &    536 &     20 &       8 &            2.800 &    10720 &                  0.000 &              1.000 \\
   timss &    757 &     23 &      13 &            1.652 &    17411 &                  0.000 &              1.000 \\
        ecpe &   2922 &     28 &       3 &            1.321 &    81816 &                  0.000 &              1.000 \\
 assistments &   4217 &  26688 &     123 &            0.796 &   346860 &                  0.997 &              1.014 \\
    berkeley &   1730 &    234 &      29 &            1.000 &   562201 &                  0.269 &              1.901 \\
    castor &  58939 &     17 &       2 &            1.471 &  1001963 &                  0.000 &              1.000 \\
\bottomrule
\end{tabular}

\label{datasets}
\end{table*}

\subsection{Relation to existing models}

When \(\psi = \logit\), KTMs include IRT, AFM and PFA. Let us now
recover some particular cases, especially when \(d = 0\), i.e., only
biases are learned for features, no embeddings. We will again assume
there are \(n\) students, \(m\) questions and \(s\) skills.

We will note \(\one_{i, n}\) a one-hot vector of size \(n\), which means
all its components are 0 except the \(i\)th one, which is 1.

\subsubsection{Relation to IRT}

If \(d = 0\), the second sum in Equation \ref{ktm} disappears and all
that is left is a weighted sum of biases.

If all features considered are students and questions (encoding users +
items), and the encoding of the pair (student \(i\), question \(j\)) is
a concatenation of one-hot vectors \(\one_{i, n}\) and \(\one_{j, m}\):
\(N = n + m\) and \(x_k = 1\) iff \(k = i\) or \(k = n + j\). The
expression in Equation \ref{ktm} becomes:

\[ \log \frac{p(\x)}{1 - p(\x)} = \mu + w_i + w_{n + j} = \theta_i - d_j \]

\noindent if the first \(n\) features (students numbered \(i\) where
\(1 \leq i \leq n\)) have bias \(w_i = \theta_i - \mu\) and the next
\(m\) features (questions numbered \(n + j\) where \(1 \leq j \leq m\))
have bias \(-d_j\). Therefore, KTM becomes after reparametrization
\(\bm{w} = (\theta_1 - \mu, \ldots, \theta_n - \mu, -d_1, \ldots, -d_m)\)
the 1-PL IRT model, also referred to as Rasch model.

\subsubsection{Relation to AFM and PFA}

Now we will again consider the special case \(d = 0\) and an encoding of
skills, wins and fails at skill level. For this, we will assume we know
the q-matrix, that is, the binary mapping between questions and skills
\((q_{jk})_{1 \leq j \leq m, 1 \leq k \leq s}\) as described in the
introduction.

If
\(\bm{w} = (\beta_1, \ldots, \beta_s, \gamma_1, \ldots, \gamma_s, \delta_1, \ldots, \delta_s)\)
and encoding of ``student \(i\) attempted question \(j\)'' is
\(\bm{x} = (q_{j1}, \ldots,\allowbreak q_{js}, W_{i1}, \ldots, W_{is}, F_{i1}, \ldots, F_{is})\),
where \(W_{ik}\) and \(F_{ik}\) are the counters of successful and
unsuccessful attempts at skill level, then KTM behaves like the PFA
model. Similarly, one can recover the AFM model.

\subsubsection{Relation to MIRT}

If \(d > 0\), KTM becomes a MIRT model with user bias:

\[ \logit p(x) = \mu + \theta_i - d_j + \langle \bm{\theta_i}, \bm{d_j} \rangle \]

The encoding is the same as for IRT (users + items with one-hot
vectors), and the embeddings
\(\bm{V} = (\bm{\theta_1}, \ldots, \bm{\theta_n}, \bm{d_1}, \ldots, \bm{d_m})\).

\subsection{Training}

Training of KTMs is made by minimizing the negative log-likelihood over
the \(S\) observations \(\bm{X} = (\x_i)_{1 \leq i \leq S}\) and
outcomes \(\bm{y} = (y_i)_{1 \leq i \leq S} \in \{0, 1\}^S\):

\[ NLL(p(\bm{X}), \bm{y}) = \sum_{i = 1}^S y_i \log p(\x_i) + (1 - y_i) \log (1 - p(\x_i)). \]

Like \cite{rendle2012factorization}, we assume some priors over the
model parameters in order to guide training and avoid overfitting.

Each bias \(w_k\) follows \(w_k \sim \N(\mu, 1/\lambda)\) and each
embedding component \(v_{kf}, f = 1, \ldots, d\) also follows
\(v_{kf} \sim \N(\mu, 1/\lambda)\) where \(\mu\) and \(\lambda\) are
regularization parameters that follow hyperpriors \(\mu \sim \N(0, 1)\)
and \(\lambda \sim \Gamma(1, 1)\).

Because of those hyperpriors, we do not need to tune regularization
parameters by hand \cite{rendle2012factorization}. As we use
\(\psi = \textnormal{probit}\), that is, the inverse of the CDF of the
normal distribution, we can fit the model using Gibbs sampling. Details
of the computations can be found in \cite{freudenthaler2011bayesian}.

The model is learned using the MCMC Gibbs sampler implementation of
libFM\footnote{\url{http://libfm.org}} in C++
\cite{rendle2012factorization}, using the pywFM Python
wrapper\footnote{\url{https://github.com/jfloff/pywFM}}.

\subsection{Visualizing the embeddings}

\begin{figure}
\includegraphics[width=\linewidth]{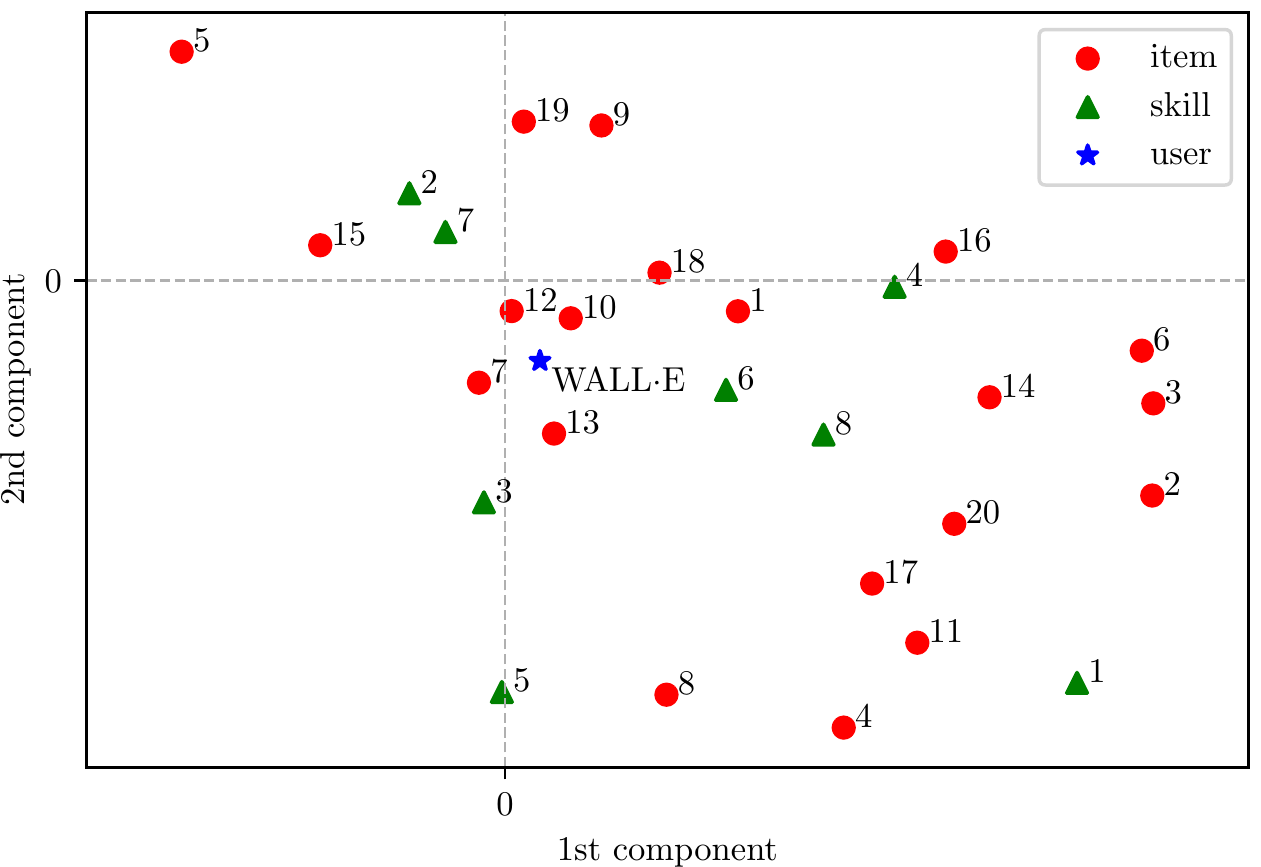}
\caption{Example of learned 2-dimensional embeddings for the Fraction dataset.}
\label{viz}
\end{figure}

Another advantage of KTMs is that we can visualize the embeddings that
they learn. On Figure~\ref{viz}, we show the 2-dimensional embeddings of
users, items, skills learned by a knowledge tracing machine on the
Fraction subtraction dataset. The user WALL·E is positively correlated
with most of items, but not skills 2 (separate a whole number from a
fraction) and 7 (substract numerators), which may explain why WALL·E
couldn't solve item 5 (\(4~3/5 - 3~4/10\)) that requires these two
skills. To know more about the items and skills of this dataset, see
\cite{DeCarlo2010}.

\section{Experiments}

We used various datasets of different shapes and sizes in order to push
our method to its limits. In Table \ref{datasets}, we report the main
characteristics of the datasets: number of users, number of items,
number of skills, average number of skills per item, total number of
observed entries, sparsity of the (user, item) pairs, average number of
attempts per user at item level.

\subsection{Temporal Datasets}

For the temporal datasets, students could attempt several times a same
question, and potentially learn between attempts.

\subsubsection{Assistments}

The 2009--2010 dataset of Assistments described in
\cite{feng2009addressing}. 4217 students over 26688 questions, 123 KCs.
347k observations. There are many items but they involve 0 to 4 KCs, and
there are only 146 combinations of KCs. For this dataset, we had also
access to more side information, referred to as ``extra'' in the
experiments:

\begin{itemize}
\tightlist
\item
  \texttt{first\_action}: attempt, or ask for a hint;
\item
  \texttt{school\_id} where the problem was assigned;
\item
  \texttt{teacher\_id} who assigned the problem;
\item
  \texttt{tutor\_mode}: tutor, test mode, pretest, or posttest.
\end{itemize}

\subsubsection{Berkeley}

1730 students from Berkeley attempting 234 questions from an online CS
course, 29 KCs, exactly 1 KC per question, which is actually a category.
650k entries.

\subsection{Non-temporal Datasets}

For all these datasets, the observations are fully specified: all users
attempted all questions. All datasets except Castor6e can be found in
the R package CDM \cite{george2016r}.

\subsubsection{Castor}

58939 middle-school students over CS-related 17 tasks, 2 KCs, 1.47 KCs
per task. 1M entries.

\subsubsection{ECPE}

2922 students over 28 language-related items, 3 KCs, 1.3 KCs per
question in average. 81k entries. This dataset can be found in the CDM R
package.

\subsubsection{Fraction}

536 middle-school students over 20 fraction subtraction questions, 8
KCs, 2.8 KCs per question in average. 16k entries. A precise description
of the items and skills is in \cite{DeCarlo2010}.

\subsubsection{TIMSS}

757 students over 23 math questions from the TIMSS test in 2003, 13 KCs,
1.65 KCs per task. 17k entries.

\subsection{Framework}

From the triplets (\texttt{user\_id}, \texttt{item\_id},
\texttt{outcome}), we first compute for the temporal datasets the number
of successful and unsuccessful attempts at skill level, according to the
q-matrix.

For each dataset, we perform 5-fold cross validation. For each fold,
entries are separated into a train and test set, then we train different
encodings of KTMs using the train set, notably the ones corresponding to
existing models, and predict the outcomes in the test set.

KTMs are trained during 1000 epochs for each non-temporal dataset, 500
epochs for the Assistments dataset and 300 epochs for the Berkeley
dataset, because it was enough for convergence. At each epoch, we
average the results over all 5 folds, in terms of accuracy (ACC), area
under the curve (AUC) and negative log-likelihood (NLL).

As special cases, as shown earlier, we have, for the temporal datasets:

\begin{itemize}
\tightlist
\item
  AFM is actually ``skills, attempts \(d = 0\)''
\item
  PFA is actually ``skills, wins, fails \(d = 0\)''
\end{itemize}

And for every dataset:

\begin{itemize}
\tightlist
\item
  IRT is ``users, items \(d = 0\)''
\item
  MIRT plus a user bias (coined as MIRTb) is ``users, items'' with any
  \(d > 0\).
\end{itemize}

\section{Results and Discussion}

\begin{figure}[t]
\includegraphics[width=\linewidth]{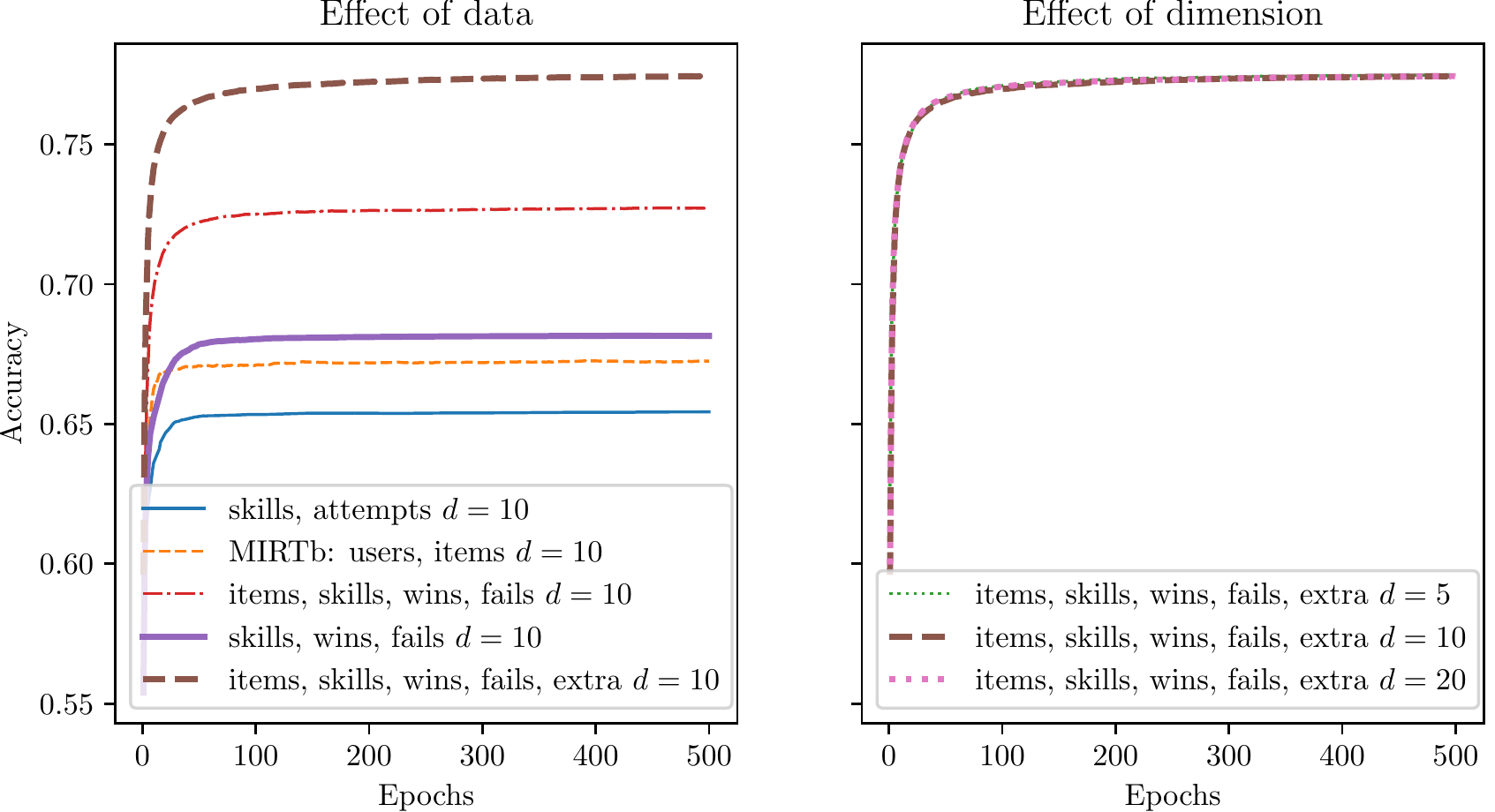}
\caption{Results for the Assistments dataset.}
\label{assistments42-figure}
\end{figure}

\begin{table}[t]
\centering
\caption{Results for the Assistments dataset.}
\footnotesize
\begin{tabular}{ccccc}
\toprule
                             model &  dim &             ACC &             AUC &             NLL \\
\midrule
 items, skills, wins, fails, extra &  20 &  \textbf{0.774} &  \textbf{0.819} &  0.465 \\
 items, skills, wins, fails, extra &  5 &  \textbf{0.775} &  \textbf{0.819} &  0.465 \\
 items, skills, wins, fails, extra &  10 &  \textbf{0.775} &  \textbf{0.818} &  0.465 \\
 items, skills, wins, fails, extra &  0 &  \textbf{0.774} &  0.815 &  \textbf{0.463} \\
 items, skills, wins, fails &  10 &  0.727 &  0.767 &  0.539 \\
 items, skills, wins, fails &  0 &  0.725 &  0.759 &  0.542 \\
 items, skills, wins, fails &  5 &  0.714 &  0.75 &  0.56 \\
 items, skills, wins, fails &  20 &  0.714 &  0.75 &  0.564 \\
 IRT: users, items &  0 &  0.675 &  0.691 &  0.599 \\
 MIRTb: users, items &  20 &  0.674 &  0.691 &  0.602 \\
 MIRTb: users, items &  10 &  0.673 &  0.687 &  0.604 \\
 MIRTb: users, items &  5 &  0.67 &  0.685 &  0.605 \\
 PFA: skills, wins, fails &  0 &  0.68 &  0.685 &  0.604 \\
 skills, wins, fails &  20 &  0.649 &  0.684 &  0.603 \\
 skills, wins, fails &  5 &  0.649 &  0.683 &  0.604 \\
 skills, wins, fails &  10 &  0.649 &  0.683 &  0.604 \\
 skills, attempts &  20 &  0.623 &  0.62 &  0.631 \\
 skills, attempts &  5 &  0.626 &  0.619 &  0.63 \\
 skills, attempts &  10 &  0.622 &  0.619 &  0.632 \\
 AFM: skills, attempts &  0 &  0.653 &  0.616 &  0.631 \\
\bottomrule
\end{tabular}

\label{assistments42-table}
\end{table}

\begin{table}[t]
\centering
\caption{Results for the Berkeley dataset.}
\footnotesize
\begin{tabular}{ccccc}
\toprule
                      model &  dim &             ACC &             AUC &             NLL \\
\midrule
 items, skills, wins, fails &  20 &  \textbf{0.706} &  \textbf{0.778} &  \textbf{0.563} \\
 items, skills, wins, fails &  10 &  \textbf{0.706} &  \textbf{0.778} &  \textbf{0.563} \\
 items, skills, wins, fails &  5 &  \textbf{0.706} &  \textbf{0.778} &  \textbf{0.563} \\
 items, skills, wins, fails &  0 &  \textbf{0.705} &  0.775 &  0.566 \\
 IRT: users, items &  0 &  0.688 &  0.753 &  0.586 \\
 MIRTb: users, items &  5 &  0.685 &  0.753 &  0.589 \\
 MIRTb: users, items &  10 &  0.685 &  0.752 &  0.59 \\
 MIRTb: users, items &  20 &  0.683 &  0.752 &  0.591 \\
 PFA: skills, wins, fails &  0 &  0.631 &  0.684 &  0.635 \\
 skills, wins, fails &  10 &  0.631 &  0.684 &  0.635 \\
 skills, wins, fails &  20 &  0.631 &  0.684 &  0.635 \\
 skills, wins, fails &  5 &  0.631 &  0.684 &  0.635 \\
 skills, attempts &  20 &  0.621 &  0.675 &  0.639 \\
 AFM: skills, attempts &  0 &  0.621 &  0.675 &  0.639 \\
 skills, attempts &  10 &  0.621 &  0.675 &  0.639 \\
 skills, attempts &  5 &  0.621 &  0.675 &  0.639 \\
\bottomrule
\end{tabular}

\label{berkeley42-table}
\end{table}

\begin{table*}[t]
\centering
\caption{Summary of AUC results for all datasets.}
\small
\begin{tabular}{lllllllll}
\toprule
{} &     AFM &     PFA &              IRT &          MIRTb10 &          MIRTb20 & KTM(iswf0) &      KTM(iswf20) & KTM(iswfe5) \\
\midrule
assistments         &  0.6163 &  0.6849 &           0.6908 &           0.6874 &           0.6907 &     0.7589 &           0.7502 &          \textbf{0.8186} \\
berkeley            &   0.675 &  0.6839 &           0.7532 &           0.7521 &           0.7519 &     0.7753 &  \textbf{0.7780} &          -- \\
ecpe                &      -- &      -- &  \textbf{0.6811} &           0.6807 &  \textbf{0.6810} &         -- &               -- &          -- \\
fraction            &      -- &      -- &           0.6662 &           0.6653 &  \textbf{0.6672} &         -- &               -- &          -- \\
timss           &      -- &      -- &  \textbf{0.6946} &           0.6939 &           0.6932 &         -- &               -- &          -- \\
castor            &      -- &      -- &  \textbf{0.7603} &  \textbf{0.7602} &           0.7599 &         -- &               -- &          -- \\
\bottomrule
\end{tabular}

\label{summary}
\end{table*}

\begin{table}[b]
\centering
\caption{Results for the Fraction dataset.}
\begin{tabular}{ccccc}
\toprule
                model &  dim &             ACC &             AUC &             NLL \\
\midrule
 MIRTb: users, items &  20 &  0.619 &  \textbf{0.667} &  0.651 \\
 items, skills &  5 &  0.621 &  \textbf{0.667} &  \textbf{0.650} \\
 items, skills &  20 &  0.621 &  \textbf{0.666} &  \textbf{0.649} \\
 MIRTb: users, items &  5 &  0.621 &  \textbf{0.666} &  \textbf{0.650} \\
 IRT: users, items &  0 &  \textbf{0.623} &  \textbf{0.666} &  0.656 \\
 users, items, skills &  0 &  \textbf{0.623} &  \textbf{0.666} &  0.656 \\
 MIRTb: users, items &  10 &  0.618 &  0.665 &  0.652 \\
 users, skills &  5 &  0.62 &  0.664 &  \textbf{0.649} \\
\bottomrule
\end{tabular}

\label{fraction42-table}
\end{table}

\begin{table}[b]
\centering
\caption{Results for the TIMSS dataset.}
\begin{tabular}{ccccc}
\toprule
                model &  dim &             ACC &             AUC &             NLL \\
\midrule
 items, skills &  0 &  0.637 &  \textbf{0.695} &  \textbf{0.629} \\
 IRT: users, items &  0 &  \textbf{0.640} &  \textbf{0.695} &  0.63 \\
 users, items, skills &  0 &  \textbf{0.639} &  \textbf{0.694} &  0.63 \\
 MIRTb: users, items &  10 &  0.638 &  \textbf{0.694} &  \textbf{0.628} \\
 MIRTb: users, items &  20 &  0.636 &  0.693 &  \textbf{0.629} \\
 users, skills &  0 &  0.579 &  0.605 &  0.67 \\
\bottomrule
\end{tabular}

\label{timss200342-table}
\end{table}

Results are reported in Tables \ref{assistments42-table} to
\ref{timss200342-table} and Figure~\ref{assistments42-figure}. For
convenience, we also reported a summary of the main results in
Table~\ref{summary}. Each existing model is matched or outperformed by a
KTM. For all non-temporal datasets, we did not consider attempt count,
as each user only attempted an item once.

\subsection{Training time}

On the Assistments dataset, for \(d = 0\), our model KTM(iswfe0) =
``items, skills, fails, extra \(d = 0\)'' is logistic regression, so it
was faster to train (4 min 30 seconds on CPU for all 5 folds) than DKT
(1 hour on CPU), while achieving higher AUC (\(0.815 > 0.743\)). For
models of higher dimensions on this dataset, experiments took 17 min for
\(d = 10\) with the same 31138 features, and 32 min for \(d = 20\).

\subsection{Effect of the side information considered}

Given its simplicity, IRT has a remarkable performance on all datasets
considered, even on the temporal ones, which may be because the average
number of attempts per student is small. When considering all
information at hand, the top performing KTM model on the Assistments
dataset for \(d = 0\) achieves higher performance than the known results
of vanilla DKT. It makes sense, as we have access to more side
information, and logistic regression is less prone to overfitting.

\paragraph{Wins and fails}

For all temporal datasets, encoding wins and fails (PFA model) instead
of only the number of attempts (AFM model) improves the performance a
lot (+0.07 AUC for Assistments, +0.01 for Berkeley). This is concordant
with existing work \cite{pavlik2009performance}. There is an improvement
of KTM models that consider number of wins and fails (KTM(iswf0) =
``items, skills, wins, fails \(d = 0\)'') over IRT (+0.07 in
Assistments, +0.02 in Berkeley).

\paragraph{Item bias}

For all datasets, considering a bias per item improves the predictions,
which is what IRT does but PFA does not. KTM(iswf0) = ``items, skills,
wins, fails \(d = 0\)'' has +0.07 AUC improvement over KTM(swf0) = PFA
in Assistments, +0.09 in Berkeley. It may be because the number of items
is huge, and they do not have the same difficulty. So, it is useful to
learn this difficulty parameter using the performance of previous
students. This extra parameter enables a big improvement on all
datasets, except on the Fraction dataset, which may be because the
skills for fraction subtraction are easily known and clearly specified,
so it is enough to characterize the items uniquely.

\paragraph{Skills}

For Fraction (8 KCs), Assistments (123 KCs) and TIMSS (13 KCs), the
skills are easy to identify, because the items are math problems. For
the other datasets, either there are few skills (ECPE: 3
language-learning KCs, Castor: 2 KCs for CS), or there is only one KC
mapped to an item (Berkeley: 29 KCs, categories of CS problems). This is
why considering a bias per skill barely increases the performance of the
predictions.

\subsection{Effect of the dimension of features}

On the temporal datasets, there is only a slight improvement of models
with higher dimensions (less than +0.01 AUC), which seems to indicate
that when there are many features considered (number of successful and
unsuccessful attempts at item or skill level), a KTM with \(d = 0\)
provides good enough predictions. Still, on a similar task,
\cite{Vie2018} managed to get an improvement of +0.03 AUC for
factorization machines for \(d = 20\) compared to logistic regression
(\(d = 0\)), presumably because the side information was considerable
for this task.

\section{Further work}

In this work, we wanted to compare the expressiveness of models
typically used for student modeling. Our experiments assess the strong
generalization of student models, as students are randomly shuffled into
train and test set, and the task of performance prediction is made for
totally new students.

\subsection{Side information in deep knowledge tracing}

The vanilla DKT model cannot handle multiple skills, so instead,
practictioners treat combinations of skills as new skills, which
prevents the transfer of information between skills. The approach
described in this paper can be used to handle multiple skills with DKT.
Also, more recent results have successfully built upon the vanilla DKT
(AUC 0.91 \textgreater{} 0.743), by incorporating dynamic cluster
information \cite{Minn2018}. We could indeed combine DKT with side
information.

\subsection{Adaptive testing}

IRT and MIRT were initially designed to provide adaptive testing: choose
the best next question to present to a learner, given their previous
answers. KTMs could also be used to these ends, as they extend the IRT
and MIRT models with extra information, under the form of KCs or several
attempts.

\subsection{Response time, spaced repetition, and other data}

Modeling response time could provide better predictions of outcomes, and
it has also been used in the encoding of factorization machines in
previous works. Also, we could add to the side information another
counter representing how many timesteps were elapsed since a certain
item was asked for the last time. It would learn how the user reacts to
spaced repetition. In some datasets such as Assistments, more data is
recorded about students that can be used to improve the predictions.
Still, we should be careful about encoding noisy data such as the output
of other machine-learning algorithms as side information, because it may
degrade performance \cite{Vie2018}.

\subsection{Higher order factorization machines}

In this paper, we were limited to pairwise interactions. But in his
original paper (\citeyear{rendle2012factorization}),
\citeauthor{rendle2012factorization} mentions higher-order factorization
machines, which generalize interactions to \(k\)-way terms. It could be
an interesting direction for future research.

\subsection{Ordinal regression}

Instead of binary outcomes, one could consider graded outcomes, with the
same KTM model, using thresholds, just like the graded response model in
item response theory \cite{samejima1997graded}. We leave it to further
work.

\section{Conclusion}

In this paper, we showed how knowledge tracing machines, a family of
models that encompasses existing models in the EDM literature as special
cases, could be used for the classification problem of knowledge
tracing.

We showed, using many datasets of various sizes and characteristics,
that it could estimate user and item parameters even when the
observations are sparse, and provide better predictions than existing
models, including deep neural networks. KTMs are a testbed to try new
combinations of data, such as response time, of number of attempts at
item level.

One can refine the encoding of features in a KTM according to how the
data was collected: Are the observations made at skill level or problem
level? Does it make sense to count the number of attempts at item level
or at skill level? What are extra sources of information that may raise
better understanding of the observations?

Furthermore, as we showed, KTMs are log-bilinear models, so the
embeddings they learn are interpretable, and can be used to provide
useful feedback to students.

\section{Acknowledgements}

We thank Mohammad Emtiyaz Khan, and the reviewers, for their precious
comments. We also thank Armando Fox and Nikunj Jain for providing the
Berkeley dataset and Mathias Hiron for providing the Castor dataset.
Part of this research was discovered in a plane, so we also thank the
flight attendants, that are always working hard to ensure our comfort.

\bibliography{biblio.bib}
\bibliographystyle{aaai}

\end{document}